\documentclass[aps,showpacs,twocolumn]{revtex4}
\usepackage{graphicx}

\begin{document}

\title{Testing Ho\v{r}ava-Lifshitz gravity using thin accretion disk
properties}
\author{Tiberiu Harko}
\email{harko@hkucc.hku.hk}
\affiliation{Department of Physics and Center for Theoretical
and Computational Physics,
The University of Hong Kong, Pok Fu Lam Road, Hong Kong}
\author{Zolt\'{a}n Kov\'{a}cs}
\affiliation{Department of Physics and Center for Theoretical
and Computational Physics,
The University of Hong Kong, Pok Fu Lam Road, Hong Kong}
\author{Francisco S. N. Lobo}
\email{flobo@cii.fis.ul.pt} \affiliation{Centro de F\'{i}sica
Te\'{o}rica e Computacional, Faculdade de Ci\^{e}ncias da
Universidade de Lisboa, Avenida Professor Gama Pinto 2, P-1649-003
Lisboa, Portugal}
\date{\today}

\begin{abstract}

Recently, a renormalizable gravity theory with higher spatial
derivatives in four dimensions was proposed by Ho\v{r}ava. The
theory reduces to Einstein gravity with a non-vanishing
cosmological constant in IR, but it has improved UV behaviors. The
spherically symmetric black hole solutions for an arbitrary
cosmological constant, which represent the generalization of the
standard Schwarzschild-(A)dS solution, has also been obtained for
the Ho\v{r}ava-Lifshitz theory. The exact asymptotically flat
Schwarzschild type solution of the gravitational field equations
in Ho\v{r}ava gravity contains a quadratic increasing term, as
well as the square root of a fourth order polynomial in the radial
coordinate, and it depends on one arbitrary integration constant.
The IR modified Ho\v{r}ava gravity seems to be consistent with the
current observational data, but in order to test its viability
more observational constraints are necessary. In the present paper
we consider the possibility of observationally testing Ho\v{r}ava
gravity by using the accretion disk properties around black holes.
The energy flux, temperature distribution, the emission spectrum
as well as the energy conversion efficiency are obtained, and
compared to the standard general relativistic case. Particular
signatures can appear in the electromagnetic spectrum, thus
leading to the possibility of directly testing Ho\v{r}ava gravity
models by using astrophysical observations of the emission spectra
from accretion disks.

\end{abstract}

\maketitle

%\preprint{gr-qc/yymmnnn}

\section{Introduction}

Recently, Ho\v{r}ava proposed a renormalizable gravity theory in
four dimensions which reduces to Einstein gravity with a
non-vanishing cosmological constant in IR but with improved UV
behaviors \cite{Horava:2008ih,Horava:2009uw}. The latter theory
admits a Lifshitz scale-invariance in time and space, exhibiting a
broken Lorentz symmetry at short scales, while at large distances
higher derivative terms do not contribute, and the theory reduces
to standard general relativity (GR). Since then various properties
and characteristics of the Ho\v{r}ava gravities have been
extensively analyzed, ranging from formal developments
\cite{formal}, cosmology \cite{cosmology}, dark energy
\cite{darkenergy} and dark matter \cite{darkmatter}, and
spherically symmetric solutions
\cite{BHsolutions,Park:2009zra,Lu:2009em,Kehagias:2009is}.
Although a generic vacuum of the theory is anti-de Sitter one,
particular limits of the theory allow for the Minkowski vacuum. In
this limit post-Newtonian coefficients coincide with those of the
pure GR. Thus, the deviations from the conventional GR can be tested
only beyond the post-Newtonian corrections, that is for a system
with strong gravity at astrophysical scales.

In this context, IR-modified Ho\v{r}ava gravity seems to be
consistent with the current observational data, but in order to
test its viability more observational constraints are necessary.
In Ref.~\cite{Konoplya:2009ig}, potentially observable properties
of black holes in the Ho\v{r}ava-Lifshitz gravity with Minkowski
vacuum were considered, namely, the gravitational lensing and
quasinormal modes. It was shown that the bending angle is
seemingly smaller in the considered Ho\v{r}ava-Lifshitz gravity
than in GR, and the quasinormal modes of black holes are longer
lived, and have larger real oscillation frequency in the
Ho\v{r}ava-Lifshitz gravity than in GR. In Ref.~\cite{Chen:2009eu},
by adopting the strong field limit approach,
the properties of strong field gravitational lensing in the
deformed H\v{o}rava-Lifshitz black hole were considered, and the
angular position and magnification of the relativistic images were
obtained. Compared with the Reissner-Norstr\"{om} black hole,
a significant difference in the parameters was found. Thus, it
was argued this may offer a way to distinguish a deformed
H\v{o}rava-Lifshitz black hole from a Reissner-Norstr\"{om} black
hole. In Ref.~\cite{Chen:2009bu}, the behavior of the effective
potential was analyzed, and the timelike geodesic motion in the
Ho\v{r}ava-Lifshitz spacetime was also explored. In this paper, we
further explore the possibility of testing the viability of
Ho\v{r}ava-Lifshitz gravity using thin accretion disk properties.

Recent observations suggest that around almost all of the active
galactic nuclei (AGN's), or black hole candidates, there exist gas
clouds surrounding the central compact object, together with an
associated accretion disc, on a variety of scales from a tenth of
a parsec to a few hundred parsecs \cite{UrPa95}. These gas clouds
are assumed to form a geometrically and optically thick torus (or
warped disc), which absorbs most of the ultraviolet radiation and
the soft X-rays. The gas exists in either the molecular or the
atomic phase. The most powerful evidence for the existence of
super massive black holes comes from the VLBI imaging of molecular
$\mathrm{H_{2}O}$ masers in the active galaxy NGC 4258 \cite
{Miyo95}. This imaging, produced by Doppler shift measurements
assuming Keplerian motion of the masering source, has allowed a
quite accurate estimation of the central mass, which has been
found to be a $3.6\times 10^{7}M_{\odot }$ super massive dark
object, within $0.13$ parsecs. Hence, important astrophysical
information can be obtained from the observation of the motion of
the gas streams in the gravitational field of compact objects.

The mass accretion around rotating black holes was studied in
general relativity for the first time in \cite{NoTh73}.  By using
an equatorial approximation to the stationary and axisymmetric
space-time of rotating black holes, steady-state thin disk models
were constructed, extending the theory of non-relativistic
accretion \cite{ShSu73}. In these models hydrodynamical
equilibrium is maintained by efficient cooling mechanisms via
radiation transport, and the accreting matter has a Keplerian
rotation. The radiation emitted by the disk surface was also
studied under the assumption that black body radiation would
emerge from the disk in thermodynamical equilibrium. The radiation
properties of the thin accretion disks were further analyzed  in
 \cite{PaTh74} and in \cite{Th74}, where the effects of the photon
capture by the hole on the spin evolution were presented as well.
In these works the efficiency with which black holes convert rest
mass into outgoing radiation in the accretion process was also
computed.

More recently, the emissivity properties of the accretion disks
were investigated for exotic central objects, such as wormholes
\cite{Harko:2008vy,HaKoLo09}, and non-rotating or rotating quark,
boson or fermion stars, brane-world black holes or gravastars
\cite{Bom,To02,YuNaRe04,Guzman:2005bs,Pun:2008ua,KoChHa09, grav}.
The radiation power per unit area, the temperature of the disk and
the spectrum of the emitted radiation were given, and compared
with the case of a Schwarzschild black hole of an equal mass. The
physical properties of matter forming a thin accretion disk in the
static and spherically symmetric spacetime metric of vacuum $f(R)$
modified gravity models were also analyzed \cite{Pun:2008ae}.
Particular signatures can appear in the electromagnetic spectrum,
thus leading to the possibility of directly testing modified
gravity models by using astrophysical observations of the emission
spectra from accretion disks.

It is the purpose of the present paper to study the thin accretion
disk models applied for black holes in Ho\v{r}ava-Lifshitz gravity
models, and carry out an analysis of the properties of the
radiation emerging from the surface of the disk. As compared to
the standard general relativistic case, significant differences
appear in the energy flux and electromagnetic spectrum for
Ho\v{r}ava black holes, thus leading to the possibility of
directly testing the Ho\v{r}ava-Lifshitz theory by using
astrophysical observations of the emission spectra from accretion
disks.

The present paper is organized as follows. In Sec. \ref{sec:II},
we present the action and specific solutions of static and
spherically symmetric spacetimes. In Sec. \ref{sec:III}, we review
the formalism and the physical properties of the thin disk
accretion onto compact objects. In Sec. \ref{sec:IV}, we analyze
the basic properties of matter forming a thin accretion disk
around vacuum black holes in Ho\v{r}ava gravity, and compare the
results with the Schwarzschild solution. We discuss and conclude
our results in Sec. \ref{sec:concl}.

\section{Black holes in Ho\v{r}ava gravity}
\label{sec:II}

In this section, we briefly review the Ho\v{r}ava-Lifshitz theory,
where differential geometry of foliations represents the proper
mathematical setting for the class of gravity theories studied by
Ho\v{r}ava \cite{Horava:2009uw}. As foliations can be equipped
with a Riemannian structure, the dynamical variables in
Ho\v{r}ava-Lifshitz gravity is the lapse function, $N$, the shift
vector $N^i$, and the 3-dimensional spatial metric, $g_{ij}$.
Thus, it is useful to use the ADM formalism, where the
four-dimensional metric is parameterized by the following
\begin{equation}
ds^2=-N^2c^2\,dt^2+g_{ij}\left(dx^i+N^i\,dt\right)
\left(dx^j+N^j\,dt\right).
\end{equation}
In this context, the Einstein-Hilbert action is given by
\begin{equation}
S=\frac{1}{16\pi G}\int d^4x \;\sqrt{g}\,N\left(K_{ij}K^{ij}-
K^2+R^{(3)}-2\Lambda \right) , \label{EHaction}
\end{equation}
where $G$ is Newton's constant, $R^{(3)}$ is the three-dimensional
curvature scalar for $g_{ij}$. The extrinsic curvature, $K_{ij}$,
is defined as
\begin{equation}
K_{ij}=\frac{1}{2N}\left(\dot{g}_{ij}-\nabla_iN_j-\nabla_jN_i\right),
\end{equation}
where the dot denotes a derivative with respect to $t$, and
$\nabla_i$ is the covariant derivative with respect to the spatial
metric $g_{ij}$.

The IR-modified Ho\v{r}ava action is given by
\begin{eqnarray}
S&=&\int dt\,d^3x
\;\sqrt{g}\,N\Bigg[\frac{2}{\kappa^2}\left(K_{ij}K^{ij}-\lambda
K^2\right) -\frac{\kappa^2}{2\nu^4} C_{ij}C^{ij}
     \nonumber  \\
&&+\frac{\kappa^2\mu}{2\nu^2}\epsilon^{ijk}R^{(3)}_{il}\nabla_j
R^{(3)l}{}_{k}-\frac{\kappa^2\mu^2}{8}R^{(3)}_{ij}R^{(3)ij}
    \nonumber  \\
&&    +\frac{\kappa^2\mu^2} {8(3\lambda-1)}
\left(\frac{4\lambda-1}{4}(R^{(3)})^2-\Lambda_W
R^{(3)}+3\Lambda_W^2\right)
    \nonumber   \\
&& +\frac{\kappa^2\mu^2\omega}{8(3\lambda-1)} R^{(3)}\Bigg],
\label{Haction}
\end{eqnarray}
where $\kappa$, $\lambda$, $\nu$, $\mu$, $\omega$ and $\Lambda_W$
are constant parameters. $C^{ij}$ is the Cotton tensor, defined as
\begin{equation}
C^{ij}=\epsilon^{ikl}\nabla_k\left(R^{(3)j}{}_{l}-\frac{1}{4}
R^{(3)}\delta^j_{l}\right).
\end{equation}
Note that the last term in Eq.~(\ref{Haction}) represents a `soft'
violation of the `detailed balance' condition, which modifies the
IR behavior. This IR modification term, $\mu^4 R^{(3)}$,
generalizes the original Ho\v{r}ava model (we have used the
notation of Ref. \cite{Park:2009zra}). Note that now these
solutions with an arbitrary cosmological constant represent the
analogs of the standard Schwarzschild-(A)dS solutions, which were
absent in the original Ho\v{r}ava model \cite{Park:2009zra}.

The fundamental constants of the speed of light $c$, Newton's
constant $G$, and the cosmological constant $\Lambda$ are defined
as
\begin{equation}
c^2=\frac{\kappa^2\mu^2|\lambda_W|}{8(3\lambda-1)^2}\quad
G=\frac{\kappa^2c^2}{16\pi(3\lambda-1)}\quad
\Lambda=\frac{3}{2}\Lambda_W c^2.
\end{equation}

%\subsection{Static and spherically symmetric black holes}
Throughout this work, we consider the static and spherically
symmetric metric given by
\begin{equation}
ds^2=-N^2(r)\,dt^2+\frac{dr^2}{f(r)}+r^2 \,(d\theta ^2+\sin
^2{\theta} \, d\phi ^2) \label{SSSmetric},
\end{equation}
where $N(r)$ and $f(r)$ are arbitrary functions of the radial
coordinate, $r$.

Imposing the specific case of $\lambda=1$, which reduces to the
Einstein-Hilbert action in the IR limit, one obtains the following
solution of the vacuum field equations in Ho\v{r}ava gravity,
\begin{equation}
N^2=f=1+(\omega-\Lambda_W)r^2-\sqrt{r[\omega(\omega-2\Lambda_W)r^3
+\beta]},
  \label{gensolution}
\end{equation}
where $\beta$ is an integration constant \cite{Park:2009zra}.

By considering $\beta=-\alpha^2/\Lambda_W$ and $\omega=0$ the
solution given by Eq.~(\ref{gensolution}) reduces to the Lu, Mei
and Pope (LMP) solution \cite{Lu:2009em}, given by
\begin{equation}
f=1-\Lambda_Wr^2-\frac{\alpha}{\sqrt{-\Lambda_W}}\sqrt{r}.
  \label{LMPsolution}
\end{equation}

Alternatively, considering now $\beta=4\omega M$ and
$\Lambda_W=0$, one obtains the Kehagias and Sfetsos's (KS)
asymptotically flat solution \cite{Kehagias:2009is}, given by
\begin{equation}
f=1+\omega r^2-\sqrt{r(\omega^2 r^3+4\omega M)}\,,
  \label{KSsolution}
\end{equation}
which is the only asymptotically flat solution in the family of
solutions (\ref{gensolution}). We shall use the Kehagias-Sfetsos
solution for analyzing the accretion disk properties. Note that
there is an outer (event) horizon, and an inner (Cauchy) horizon
at
\begin{equation}
r_{\pm}=M\left[1\pm\sqrt{1-1/(2\omega M^2)}\right].
\end{equation}
To avoid a naked singularity at the origin, one also needs to
impose the condition
\begin{equation}
\omega M^2\geq \frac{1}{2}.
\end{equation}
Note that in the GR regime, i.e., $\omega M^2 \gg 1$, the outer
horizon approaches the Schwarzschild horizon, $r_+\simeq 2M$, and
the inner horizon approaches the central singularity, $r_- \simeq
0$.

\section{Electromagnetic radiation properties of thin accretion disks}
\label{sec:III}

To set the stage, we present the general formalism of
electromagnetic radiation properties of thin accretion disks in a
general static, spherically-symmetric spacetime.

\subsection{Spacetime metric and geodesic equations}

In this work we analyze the physical properties and
characteristics of particles moving in circular orbits around
general relativistic compact spheres in a static and spherically
symmetric geometry given by the following metric
\begin{equation}  \label{rotmetr1}
ds^2=g_{tt}\,dt^2+g_{rr}\,dr^2
+g_{\theta\theta}\,d\theta^2+g_{\phi\phi}\,d\phi^2.
\end{equation}
Here the metric components $g_{tt}$,  $g_{rr}$, $
g_{\theta\theta}$ and $g_{\phi\phi}$ depend only on the radial
coordinate $r$. In  a static and spherically symmetric spacetime
two constants of motion for particles do exist, the specific
energy $ \widetilde{E}$ and of the specific angular momentum
$\widetilde{L}$, respectively. The geodesic equations of motion in
the equatorial plane ($\theta=\pi/2$) can be written in terms of
these constants of motion as
\begin{eqnarray}
g_{tt}\dot{t}&=&-\widetilde{E}
\,,  \label{geodeqs1} \\
g_{\phi\phi}\dot{\phi}&=&\widetilde{L}
\,,  \label{geodeqs2} \\
-g_{tt}g_{rr}\dot{r}^2+V_{eff}(r)&=&\widetilde{E}^2.
\label{geodeqs3}
\end{eqnarray}
where the effective potential term is defined as
\begin{equation}  \label{roteffpot}
V_{eff}(r)= -g_{tt}\left(1+\frac{\widetilde{L}^2}{g_{\phi\phi}}\right).
\end{equation}

For stable circular orbits in the equatorial plane the following
conditions must hold: $V_{eff}(r)=0$ and $V_{eff,\;r}(r)=0$,
respectively. These conditions provide the specific energy, the
specific angular momentum and the angular velocity $\Omega$ of
particles moving in circular orbits for the case of static general
relativistic compact spheres, given by
\begin{eqnarray}
\widetilde{E}&=&-\frac{g_{tt}}{\sqrt{-g_{tt}
-g_{\phi\phi}\Omega^2}}\,,  \label{rotE} \\
\widetilde{L}&=&\frac{g_{\phi\phi}\Omega}{\sqrt{-g_{tt}
-g_{\phi\phi}\Omega^2}}\,,  \label{rotL} \\
\Omega&=&\frac{d\phi}{dt}=\sqrt{\frac{-g_{tt,r}}{g_{\phi\phi,r}}}.
\label{rotOmega}
\end{eqnarray}
The marginally stable orbit around the central object can be
determined from the condition $V_{eff,\;rr}(r)=0$. This condition
provides the following relation
\begin{eqnarray}
\widetilde{E}^{2}g_{\phi\phi,rr}+ \widetilde{L}^{2}g_{tt,rr} +
(g_{tt}g_{\phi\phi})_{,rr} =0.  \label{mso-r}
\end{eqnarray}

By inserting Eqs. (\ref{rotE})-(\ref{rotOmega}) into
Eq.~(\ref{mso-r}), and solving this equation for $r$, we obtain
the marginally stable orbit for the explicitly given metric
coefficients $g_{tt}$, $g_{t\phi}$ and $g_{\phi\phi}$. For a
Schwarzschild black hole we have $g_{tt}=-(1-2M/r)$,
$g_{rr}=-{g}^{-1}_{tt}$ and $g_{\phi\phi}=r^2$,  and the geodesic
equation (\ref{geodeqs3}) for the radial coordinate $r$ becomes
\begin{equation}
\dot{r}^{2}+V_{eff}(r)=\widetilde{E}^2
\end{equation}
with the effective potential given by
\begin{equation}
V_{eff}(r)= \left(1-\frac{2M}{r}\right)
\left(1+\frac{\widetilde{L}^2}{r^2}\right).
\end{equation}

\begin{figure}[h]
\centering
  \includegraphics[width=3.2in]{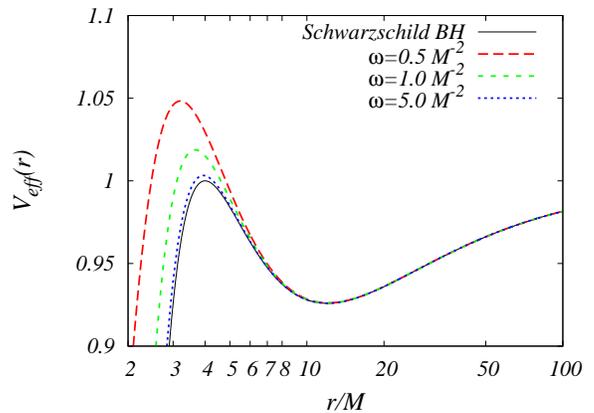}
  \caption{The effective potential $V_{eff}(r)$ of the
  orbiting particles for the Kehagias-Sfetsos solution and for the
  Schwarzschild black hole with the same total mass $M$
  for the specific angular  momentum $\widetilde{L}=4M$.
  The parameter $\omega$ of the Kehagias-Sfetsos solution is set to $0.5M^{-2}$,
  $1M^{-2}$ and $5M^{-2}$, respectively.}
  \label{Fig:effpot}
\end{figure}

Eqs.~(\ref{rotE})-(\ref{rotOmega}) leads to the form
\begin{eqnarray}
\widetilde{E}&=&\frac{r(r-2M)}{\sqrt{r-
3M}}\,,  \label{rotESch} \\
\widetilde{L}&=&\frac{R^5\Omega}{\sqrt{r-3M
}},  \label{rotLSch} \\
\Omega&=&\sqrt{\frac{M}{r^3}}.  \label{rotOmegaSch}
\end{eqnarray}
for the specific energy, the
specific angular momentum, and the angular velocity for the Schwarzschild metric. Since for the KS solution, given
by Eq.~(\ref{KSsolution}), $g_{tt}=-f(r)$, $g_{rr}=-g_{tt}^{-1}$
and $g_{\phi\phi}=r^2$, the effective potential in
Ho\v{r}ava-Lifshitz theory can be written as
\begin{equation}
V_{eff}(r)= \left[1+\omega r^2-f_{M,\omega}(r)\right]
\left(1+\frac{\widetilde{L}^2}{r^2}\right),
\end{equation}
with $f_{M,\omega}(r)\equiv\sqrt{r\omega(\omega r^3+4M)}$, whereas
the specific energy, the specific angular momentum and the angular
velocity are given by
\begin{eqnarray}
\widetilde{E}&=&\frac{\sqrt{1
+r^2\omega-f_{M,\omega}}}{\sqrt{1
+r^2(\omega-\Omega^2)-f_{M,\omega}}},  \label{rotEKS} \\
\widetilde{L}&=&\frac{r^2\Omega}{\sqrt{1
+r^2(\omega-\Omega^2)-f_{M,\omega}}},  \label{rotLKS} \\
\Omega&=&\sqrt{\frac{rf_{M,\omega}-M-\omega r^3}{r\sqrt{f_{M,
\omega}}}}.  \label{rotOmegaKS}
\end{eqnarray}
The effective potentials of the Schwarzschild black hole and of
the KS solution are compared for the same geometrical mass in
Fig.~\ref{Fig:effpot}. As previously shown in
\cite{Konoplya:2009ig}, $V_{eff}(r)$ for the KS solution
approaches the Schwarzschild potential for increasing values of
$\omega $.

\subsection{Properties of thin accretion disks}

For a thin accretion disk we assume that its vertical
size is negligible, as compared to its horizontal extension, i.e,
the disk height $H$, defined by the maximum half thickness of the
disk, is always much smaller than the characteristic radius $r$ of
the disk, $H \ll r$. The thin disk is in hydrodynamical
equilibrium, and the pressure gradient and a vertical entropy
gradient in the accreting matter are negligible. The efficient
cooling via the radiation over the disk surface prevents the disk
from cumulating the heat generated by stresses and dynamical
friction. In turn, this equilibrium causes the disk to stabilize
its thin vertical size. The thin disk has an inner edge at the
marginally stable orbit of the compact object potential, and the
accreting plasma has a Keplerian motion in higher orbits.

In steady state accretion disk models, the mass accretion rate
$\dot{M}_{0}$ is assumed to be a constant that does not change
with time. The physical quantities describing the orbiting plasma
are averaged over a characteristic time scale, e.g. $\Delta t$,
over the azimuthal angle $\Delta \phi =2\pi $ for a total period
of the orbits, and over the height $H$ \cite{ShSu73,
NoTh73,PaTh74}. In the standard accretion disk theory the
integration of the total divergence of the energy-momentum tensor
of the plasma forming the disk provides the disk structure
equations. The radiation flux $F$ emitted by the surface of the
accretion disk can be derived from the conservation equations for
the mass, energy and angular momentum, respectively, and it is
expressed in terms of the specific energy, angular momentum and of
the angular velocity of the particles orbiting in the disk as
\cite{NoTh73,PaTh74},
\begin{equation}
F(r)=-\frac{\dot{M}_{0}}{4\pi \sqrt{-g}}\frac{\Omega
_{,r}}{(\widetilde{E}-\Omega
\widetilde{L})^{2}}\int_{r_{ms}}^{r}(\widetilde{E}-\Omega
\widetilde{L}) \widetilde{L}_{,r}dr,  \label{F}
\end{equation}
where $\dot{M}_0$ is the mass accretion rate, measuring the rate
at which the rest mass of the particles flows inward through the
disk with respect to the coordinate time $t$, and $r_{ms}$ is the
marginally stable orbit obtained from Eq.~(\ref{mso-r}),
respectively.

Another important characteristics of the mass accretion process is
the efficiency with which the central object converts rest mass
into outgoing radiation. This quantity is defined as the ratio of
the rate of the radiation of energy of photons escaping from the
disk surface to infinity, and the rate at which mass-energy is
transported to the central compact general relativistic object,
both measured at infinity \cite{NoTh73,PaTh74}. If all the emitted
photons can escape to infinity, the efficiency is given in terms
of the specific energy measured at the marginally stable orbit $
r_{ms}$,
\begin{equation}
\epsilon =1-\widetilde{E}_{ms}.  \label{epsilon}
\end{equation}

For Schwarzschild black holes the efficiency is about $6\%$,
whether the photon capture by the black hole is considered, or
not. Ignoring the capture of radiation by the hole, $\epsilon $ is
found to be $42\%$ for extremely rotating Kerr black holes
($a_{*}=1$), whereas with photon capture the efficiency is $40\%$
\cite{Th74}.

The accreting matter in the steady-state thin disk model is
supposed  to be in thermodynamical equilibrium. Therefore the
radiation emitted by the disk surface can be considered as a
perfect black body radiation, where the energy flux is given by
$F(r)=\sigma T^{4}(r)$ ($\sigma $ is the Stefan-Boltzmann
constant), and the observed luminosity $L\left( \nu \right)$ has a
redshifted black body spectrum \citep{To02}:
\begin{equation}
L\left( \nu \right) =4\pi d^{2}I\left( \nu \right) =\frac{8}{\pi c^2 }\cos
\gamma \int_{r_{i}}^{r_{f}}\int_0^{2\pi}\frac{\nu^{3}_e r d\phi dr }{\exp
\left( h\nu_e/T\right) -1}.\label{L}
\end{equation}

Here $d$ is the distance to the source, $I(\nu )$ is the thermal
energy flux radiated by the disk, $\gamma $ is the disk
inclination angle, and $r_{i}$ and $r_{f}$ indicate the position
of the inner and outer edge of the disk, respectively. We take
$r_{i}=r_{ms}$ and $r_{f}\rightarrow \infty $, since we expect the
flux over the disk surface vanishes at $r\rightarrow \infty $for
any kind of asymptotically flat geometry. The emitted frequency is
given by $\nu_e=\nu(1+z)$, where the redshift factor can be
written as
\begin{equation}
1+z=\frac{1+\Omega r \sin \phi \sin \gamma }{\sqrt{ -g_{tt}-2
\Omega g_{t\phi} - \Omega^2 g_{\phi\phi}}},\label{z}
\end{equation}
where we have neglected the light bending \cite{Lu79,BMT01}.

\section{Electromagnetic signatures of accretion
disks around Kehagias-Sfestos black holes}\label{sec:IV}

As a first step in the study of the accretion disk properties, we
obtain Eqs.~(\ref{rotEKS})-(\ref{rotOmegaKS}) for the specific
energy $\widetilde{E}$, the specific angular momentum
$\widetilde{L}$ and the angular velocity $\Omega$ of any particle
orbiting around a KS black holes. By inserting
Eqs.~(\ref{rotEKS})-(\ref{rotOmegaKS}) into the flux integral
Eq.~(\ref{F}), we can derive the radial profile of the emitted
photon energy flux over the whole surface the disk in the KS
potential. Eq.~(\ref{F}) is derived by integrating the
conservation laws for the mass, energy and angular momentum, which
are invariant for Ho\v{r}ava gravity, since the extra terms in the
action Eq.~(\ref{Haction}) do not give any contribution to the
total divergence of the stress energy tensor.

\begin{figure}[h]
\centering
  \includegraphics[width=3.2in]{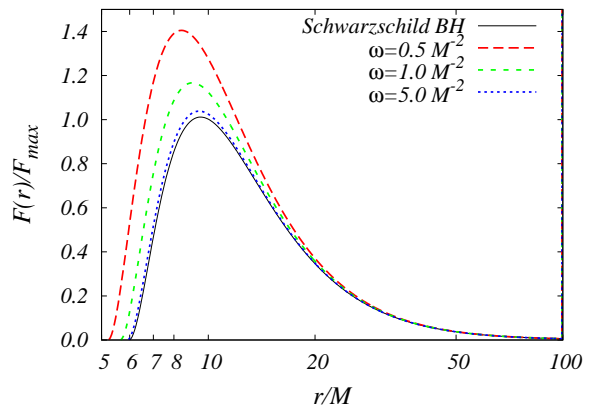}
  \caption{The energy radiated by a disk around the Kehagias-Sfetsos and
  Schwarzschild black holes with the same total mass $M$.
  The parameter $\omega$ of the Kehagias-Sfetsos solution is set to
  $0.5M^{-2}$, $1M^{-2}$ and $5M^{-2}$, respectively, and
  the flux values are normalized by
  $F_{max}=1.37\times10^{-5}\dot{M}_0/M^2$, the maximal
  flux value for the Schwarzschild black hole.}
  \label{Fig:flux}
\end{figure}

The profiles for the energy flux are presented, for different
values of $\omega$, in Fig.~\ref{Fig:flux}. For the sake of
comparison we also present the flux distribution over a disk
rotating around a Schwarzschild black hole.

Similarly to the case of the effective potential, the deviation of
$F(r)$ for the KS geometry from the standard Schwarzschild flux
increases as $\omega$ tends to $0.5M^{-2}$. The left edge of the
flux profiles, shifting from $r/M=6$ to lower radii, shows that
the distance of the inner edge of the accretion disk and the event
horizon of the KS black hole remains almost the same as for the
Schwarzschild geometry (see Table \ref{Efficiency}). For
$\omega=0.5M^{-2}$ the degenerate event horizon of the KS black
hole is at $r=M$, and the marginally stable orbit approaches
$r/M=5$. The maximal flux value also increases for smaller values
of $\omega $. When $\omega M^2$ reaches its lower limit at 0.5,
the maximum value of the flux is already a factor of 1.4 higher
than the maximum value $F_{max}=1.37\times10^{-5}\dot{M}_0/M^2$
corresponding to the Schwarzschild solution. Similarly to the
inner edge of the disk, the flux maximum is shifted to lower and
lower radii by decreasing $\omega $.

\begin{figure}[h]
\centering
  \includegraphics[width=3.2in]{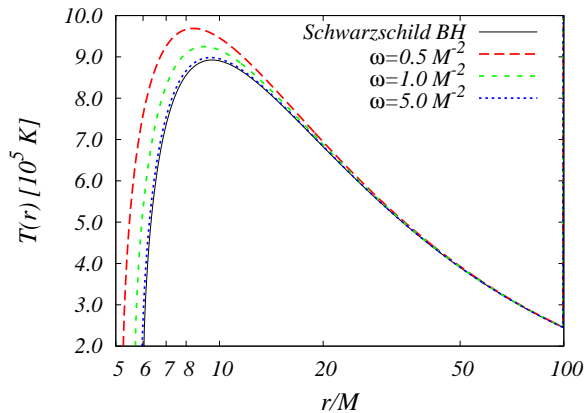}
  \caption{The disk temperature for Kehagias-Sfetsos and Schwarzschild black
  holes with the same total mass $M$. The parameter $\omega$ of
  the Kehagias-Sfetsos solution is set to $0.5M^{-2}$, $1M^{-2}$ and $5M^{-2}$,
  respectively.}
  \label{Fig:temp}
\end{figure}

These features can also be observed in the temperature profiles
presented in Fig.~\ref{Fig:temp}. However, the differences in
the temperature amplitudes are not so big as they are in the case
of the flux distribution.

\begin{figure}[h]
\centering
  \includegraphics[width=3.2in]{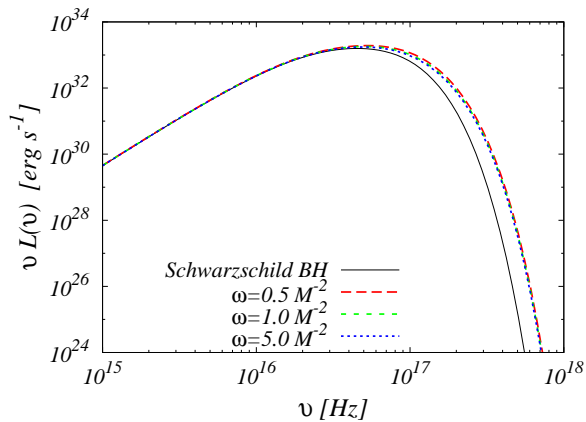}
  \caption{Disk spectra for Kehagias-Sfetsos and Schwarzschild black holes
  with the same total mass $M$. The parameter $\omega$ of
  the Kehagias-Sfetsos solution is set to $0.5M^{-2}$, $1M^{-2}$ and $5M^{-2}$,
  respectively. Here $M=1M_{\odot}$
  and $\dot{M}_0=10^{-12}M_{\odot}$/yr.}
  \label{Fig:lum}
\end{figure}

In Fig.~\ref{Fig:lum}, the spectral energy distribution,
calculated with the use of Eqs.~(\ref{L}) and (\ref{z}),
respectively, shows a more interesting difference between the disk
spectra of the KS black hole and of the Schwarzschild black hole,
respectively. The disk spectra are very similar for both the KS
and the Schwarzschild black holes in the region with $\nu\leq
10^{16}$Hz. The cut-off frequencies of the spectra are also of the
order of $\approx10^{16}$ Hz for all cases, but they are somewhat
higher for the KS black holes than for the Schwarzschild case,
which separates the two classes. For the  KS solution the spectral
properties do not exhibit any significant differences with the
variation of $\omega $: the spectra are essentially the same for
any value of $\omega$. Although the amplitude and the cut-off
frequency of the spectra are maximal in the limit
$\omega=0.5M^{-2}$, the differences in these quantities are
negligible even for $\omega=1000M^{-2}$.

Table~\ref{Efficiency} shows the conversion efficiency $\epsilon$
of the accreted mass into radiation for both KS and Schwarzschild
black holes. For a given configuration with a fixed value of
$\omega$, $\epsilon$ is somewhat higher in the accretion process
driven by KS black holes, as compared to the Schwarzschild
geometry. This means that KS black holes always convert more
efficiently mass into radiation than a standard general
relativistic, static black hole do. The most efficient mechanism
is provided by the KS black holes for the minimal value of
$\omega$, where efficiency is 6.3\%. For $\omega M^2>>1$, the
values of $\epsilon$ and $r_{ms}$ approach those of the
Schwarzschild black hole, as expected.

\begin{table}[tbp]
\begin{center}
\begin{tabular}{|c|c|c|}
\hline
$\omega$ [$M^2$] & $r_{ms}$ [$M$] & $\epsilon$ \\ \hline
0.5 & 5.2441 & 0.0630 \\ \hline
1.0 & 5.6644 & 0.0597 \\ \hline
5.0 & 5.9536 & 0.0576 \\ \hline
- & 6.0000 & 0.0572 \\ \hline
\end{tabular}%
\end{center}
\caption{The marginally stable orbit and the efficiency for Kehagias-Sfetsos and
Schwarzschild black hole geometries. The last line corresponds
to the standard
general relativistic Schwarzschild black hole.}
\label{Efficiency}
\end{table}

\section{Discussions and final remarks}
\label{sec:concl}

In the present paper we have considered the basic physical
properties of matter forming a thin accretion disc in the vacuum
spacetime metric of the Ho\v{r}ava-Lifshitz gravity models. The
physical parameters of the disc -- effective potential, flux and
emission spectrum profiles -- have been explicitly obtained for
several values of the parameter $\omega $ characterizing the
vacuum solution of the generalized field equations. All the
astrophysical quantities, related to the observable properties of
the accretion disc, can be obtained from the black hole metric.
Due to the differences in the space-time structure, the
Ho\v{r}ava-Lifshitz theory black holes present some very important
differences with respect to the disc properties, as compared to
the standard general relativistic Schwarzschild case.

The determination of the accretion rate for an astrophysical
object can give a strong evidence for the existence of a surface
of the object. A model in which Sgr A*, the $3.7\times
10^{6}M_{\odot }$ super massive black hole candidate at the
Galactic center, may be a compact object with a thermally emitting
surface was considered in \cite{BrNa06}. For very compact surfaces
within the photon orbit, the thermal assumption is likely to be a
good approximation because of the large number of rays that are
strongly gravitationally lensed back onto the surface. Given the
very low quiescent luminosity of Sgr A* in the near-infrared, the
existence of a hard surface, even in the limit in which the radius
approaches the horizon, places a severe constraint on the steady
mass accretion rate onto the source, ${\dot{M }}\leq
10^{-12}M_{\odot }$ yr$^{-1}$. This limit is well below the
minimum accretion rate needed to power the observed submillimeter
luminosity of Sgr A*, ${\dot{M}}\geq 10^{-10}M_{\odot }$
yr$^{-1}$. Thus, from the determination of the accretion rate it follows that
Sgr A* does not have a surface, that is, it must have an event
horizon.

Therefore, the study of the accretion processes by compact
objects is a powerful indicator of their physical nature. Since
the energy flux, the temperature distribution of the disk, the spectrum  of the emitted black body radiation, as well as the conversion efficiency show, in the case of the Ho\v{r}ava-Lifshitz theory vacuum solutions, significant differences as compared to the general relativistic case, the
determination of these observational quantities could discriminate, at least in
principle, between standard general relativity and Ho\v{r}ava-Lifshitz theory, and constrain
the parameter of the model.

\section*{Acknowledgments}

The work of T. H. was supported by the General Research Fund grant
number HKU 701808P of the government of the Hong Kong Special
Administrative Region.

\end{document}